\documentclass[a4paper,fleqn,usenatbib]{mnras}

\usepackage{newtxtext,newtxmath}
\usepackage[T1]{fontenc}
\usepackage{ae,aecompl}
\usepackage{graphicx}	% Including figure files
\usepackage{amsmath}	% Advanced maths commands
\usepackage{amssymb}	% Extra maths symbols
\defcitealias{arenou}{AGG}
\defcitealias{sfd}{SFD}
\defcitealias{2015ApJ...798...88M}{PLA}
\defcitealias{drimmel}{DCL}
\defcitealias{green}{GSF}
\defcitealias{av}{G12}
\defcitealias{g17}{G17}
\defcitealias{chen}{CLY}
\defcitealias{sale}{SDB}

\title[Reddening for {\it Gaia} DR1 TGAS giants]{Verifying reddening and extinction for {\it Gaia} DR1 TGAS giants}

\author[G. A. Gontcharov and A. V. Mosenkov]{
George A. Gontcharov,$^{1,2}$\thanks{E-mail: george.gontcharov@tdt.edu.vn}
and Aleksandr V. Mosenkov,$^{3,4,5}$
\\
$^{1}$Department for Management of Science and Technology Development, Ton Duc Thang University, Ho Chi Minh City, Vietnam\\
$^{2}$Faculty of Applied Sciences, Ton Duc Thang University, Ho Chi Minh City, Vietnam\\
$^{3}$Sterrenkundig Observatorium, Universiteit Gent, Krijgslaan 281, B-9000 Gent, Belgium\\
$^{4}$St.Petersburg State University, 7/9 Universitetskaya nab., St.Petersburg, 199034 Russia\\
$^{5}$Central Astronomical Observatory, Russian Academy of Sciences, 65/1 Pulkovskoye chaussee, St. Petersburg, 196140 Russia
}

\date{Accepted XXX. Received YYY; in original form ZZZ}

\pubyear{2017}

\begin{document}
\label{firstpage}
\pagerange{\pageref{firstpage}--\pageref{lastpage}}
\maketitle

\begin{abstract}
{\it Gaia} DR1 Tycho--Gaia Astrometric Solution parallaxes, {\it Tycho-2} photometry and reddening/extinction estimates from nine data sources
for 38074 giants within 415 pc from the Sun are used to compare their position in the Hertzsprung--Russell diagram with theoretical estimates, 
which are based on the PARSEC and MIST isochrones and the TRILEGAL model of the Galaxy with its parameters being widely varied.
We conclude that 
(1) some systematic errors of the reddening/extinction estimates are the main uncertainty in this study;
(2) any emission-based 2D reddening map cannot give reliable estimates of reddening within 415 pc due to a complex distribution of dust;
(3) if a TRILEGAL's set of the parameters of the Galaxy is reliable and if the solar metallicity is $Z<0.021$, then 
the reddening at high Galactic latitudes behind the dust layer is underestimated by all 2D reddening maps based on the dust emission
observations of {\it IRAS}, {\it COBE}, and {\it Planck} and by their 3D followers (we also discuss some explanations of this underestimation);
4) the reddening/extinction estimates from recent 3D reddening map by Gontcharov, 
including the median reddening $E(B-V)=0.06$~mag at $|b|>50\degr$, give the best fit of the empirical and theoretical data with each other.
\end{abstract}


\section{Introduction}
\label{intro}

In recent years, the data used for the Hertzsprung--Russell (HR) diagrams `dereddened colour versus absolute magnitude' have been greatly improved. 
For millions {\it Tycho-2} \citep{tycho2} stars we have the parallax $\varpi$ with its error $\sigma(\varpi)$ from the 
{\it Gaia} DR1 Tycho--Gaia Astrometric Solution \citep[TGAS, ][]{gaiaa, gaiab},
$(B_T-V_T)$ colour being accurate at the level of $0.05$ mag and the estimates of reddening and interstellar extinction from various data sources.
Thus, the dereddened colour $(B_T-V_T)_0$ and absolute magnitude $M_{V_T}$ can be precisely calculated as:
\begin{equation}
\label{equ0}
M_{V_T}=V_T+5-5\log(R)-A_{V_T},
\end{equation}
\begin{equation}
\label{dered}
(B_T-V_T)_0=(B_T-V_T)-E(B_T-V_T),
\end{equation}
where $R$ is the distance, $A_{V_T}$ is the extinction in $V_T$ and $E(B_T-V_T)$ is the reddening.
Other photometric bands can be also used.

The accuracy of the data is not enough for making conclusions about every star, but some statistics of the distribution of many stars in the HR diagram 
can reveal some systematic errors of the data.
Such studies have been made by \citet[][hereafter G17]{g17} and \citet{gm2017big} for O--F main-sequence stars across the sky and by
\citet{gm2017} for both O--F stars and giants in the {\it Kepler} field.
In this study, we consider the TGAS giants across the sky.
To select their sample with the best data and high level of completeness the following self-consistent constraints are applied:
$-1.5<M_{V_T}+A_{V_T}<2.5$ mag, $B_T<12$ mag, $V_T<10.5$ mag, $0.85<(B_T-V_T)<2.4$ mag, $R<415$ pc and $\sigma(\varpi)/\varpi)<0.2$.
We use $R$ from \citet{bailer3}.

\begin{figure}
\includegraphics{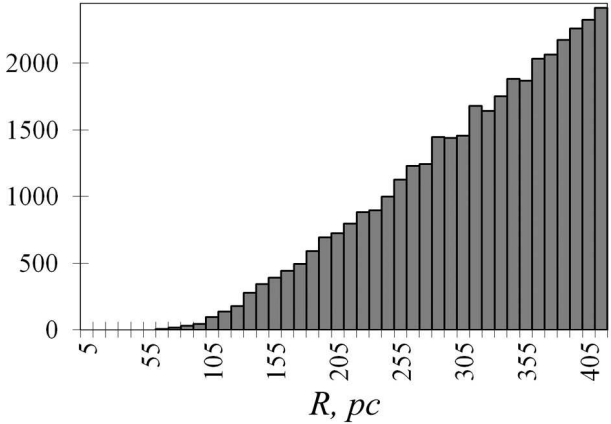}
\caption{Histogram of the distances for the giants under consideration.
}
\label{histo}
\end{figure}

Fig.~\ref{histo} shows the histogram of the distances for 38074 selected giants.
The median value of $R$ of the sample is 324 pc.
The giants are intrinsically brighter than O--F stars and, thus, can be investigated at a larger distance.
But more importantly, the giants are much more numerous far away from the Galactic mid-plane than O--F stars.
The majority of the O--F stars,which were considered by \citet{gm2017big}, are within 220 pc, thus, they are located inside of the Galactic dust layer.
On the other hand, the majority of the selected giants are within $250<R<415$ pc, thus, they are located behind the dust layer at high latitudes.
Any conclusion about the reddening through the dust layer
\footnote{In fact, it is the dust half-layer to the North or to the South due to the position of the Sun nearly in the middle of this layer.}
 at high latitudes would be very important because it defines a zero-point of many 3D reddening data sources.
We can make such a conclusion in this study.
Moreover, now we can consider the 3D reddening maps of \citet[][hereafter GSF]{green}\footnote{\url{http://argonaut.skymaps.info/}}
and \citet[][hereafter SDB]{sale}\footnote{\url{http://www.iphas.org/extinction/}}, 
which give poor data for $R<280$ pc, but reliable ones for $R<415$ pc (we consider only the giants with the reliable estimates).

Obviously, the larger $R$ the less precise the TGAS $\varpi$.
Yet, photometry in many deep surveys is saturated for close and bright TGAS stars.
Also, any infrared (IR) photometry is not enough sensitive for detection of the errors in reddening/extinction estimates.
Therefore, we choose the {\it Tycho-2} photometry as the only one, precise and sensitive enough \citep{g2016},
and available for all the TGAS stars under consideration (besides the {\it Gaia} band $G$).

Let us consider the balance of the systematic errors in the equations~(\ref{equ0}) and (\ref{dered}) for the selected sample.
A detailed study of the TGAS giants by \citet{g2017} has shown that the systematic errors for $\varpi>1.5$ mas seem to be lower than 0.1 mas.
For $R<415$ pc it gives $\sigma(M_{V_T})<0.09$~mag \citep[][ p. 44]{parenago}.
Any systematic error of $V_T$ or $(B_T-V_T)$ is less than 0.01 mag \citep{tycho2}.
\citet{gm2017big} have compared various estimates of reddenings/extinctions for the TGAS stars and found that even within $R<280$ pc
their typical systematic differences are $\Delta(E(B_T-V_T))=0.05$ and $\Delta(A_{V_T})=0.15$ mag, thus, representing their systematic errors.
Therefore, the systematic errors of reddening/extinction dominate the balance of the systematic errors in the equations~(\ref{equ0}) and (\ref{dered}).
Consequently, statistics of the distribution of the TGAS giants in the HR diagram would primarily reveal some systematic errors of the 
reddening/extinction estimates.

In this paper, for the comparison with the empirical data the proper theoretical positions of the giants in the HR diagram are calculated using 
the PARSEC \citep{bressan}\footnote{\url{http://stev.oapd.inaf.it/cgi-bin/cmd}} 
and MIST \citep{mist}\footnote{\url{http://waps.cfa.harvard.edu/MIST/}} isochrones and the
TRILEGAL model of the Galaxy \citep{trilegal}
\footnote{\url{http://stev.oapd.inaf.it/cgi-bin/trilegal}}.

\section{Data and results}
\label{data}

The distribution of the selected 38074 giants in the HR diagram $M_{V_T}$ versus $(B_T-V_T)_0$
\footnote{Hereafter, the HR diagrams are rotated by 90$\degr$ w.r.t. usual view because we consider the variations of some parameters with $M_{V_T}$.}
is shown in Fig.~\ref{hrdots1} by the grey and black dots for $|b|<50\degr$ and $|b|>50\degr$, respectively.
Different plots in Fig.~\ref{hrdots1} show different reddening/extinction estimates used:
(a) zero extinction,
(b) \citet[][hereafter AGG]{arenou},
(c) \citet[][hereafter SFD]{sfd},
(d) \citetalias{sfd} reduced as described later (hereafter SFD$_R$),
(e) \citet[][hereafter PLA]{2015ApJ...798...88M}, 
(f) \citetalias{2015ApJ...798...88M} reduced as described later (hereafter PLA$_R$).
(g) \citet[][hereafter DCL]{drimmel},
(h) \citetalias{green},
(i) \citet[][hereafter G12]{gould, av},
(j) \citetalias{g17},
(k) \citet[][hereafter CLY]{chen}\footnote{\url{http://lamost973.pku.edu.cn/site/data}} and
(l) \citetalias{sale}.
All of them, except \citetalias{green} and \citetalias{sale}, have been described and used by \citet{gm2017big} for $R<280$ pc.
\citetalias{green}, \citetalias{chen}, and \citetalias{sale} cover only parts of the sky and provide reddening/extinction estimates only for 9010, 5483 and 2143 giants under consideration, 
respectively.
\citetalias{chen} and \citetalias{sale} have no giants under consideration for $|b|>50\degr$.
The \citetalias{drimmel}, \citetalias{green} and \citetalias{sale} estimates are calculated by use of the code of \citet{bovy2016}\footnote{\url{https://github.com/jobovy/mwdust}}.

\citetalias{sfd} and \citetalias{2015ApJ...798...88M} are 2D (to infinity) reddening maps based on the observations of the dust emission in far-IR by {\it COBE}, {\it IRAS}, and {\it Planck}.
To compare them with the 3D reddening/extinction estimates we reduce them from infinity to $R$ by use of the barometric law of the dust spatial 
distribution \citep[][ p. 265]{parenago}:
\begin{equation}
\label{baro}
E(B-V)_\mathrm{R}=E(B-V)\,(1-\mathrm{e}^{-|Z-Z_0|/Z_\mathrm{A}})\,,
\end{equation}
where $E(B-V)_\mathrm{R}$ is the reddening to the distance $R$,
$E(B-V)$ is the reddening to infinity for the same line of sight,
$Z=R\sin(b)$ is the Galactic coordinate of the object along the $Z$-axis in kiloparsecs
\footnote{To avoid confusion, the metallicity is designated hereafter as $\mathbf Z$,
while one of the Galactic coordinates as $Z$.},
$Z_0$ is the vertical offset of the mid-plane of the dust layer with respect to the Sun in kiloparsecs, and
$Z_\mathrm{A}$ is the scale height of the dust layer in kiloparsecs.
Following \citet{gm2017big}, we accept some average values $Z_\mathrm{A}=100$ and $Z_0=13$ pc.
By doing so, we calculated SFD$_R$ and PLA$_R$.

Following the PARSEC data base, we use the extinction law (i.e. the dependence of extinction on wavelength) of \citet{cardelli} with the fixed
extinction-to-reddening ratio $R_V=A_V/E(B-V)=3.1$.
With enough accuracy it can be approximated for the giants with $(B_T-V_T)_0<1.9$~mag by the equations:
\begin{equation}
\label{ebtvtebv}
E(B_T-V_T)/E(B-V)=0.0295C^3-0.1139C^2+0.208C+0.96\,,
\end{equation}
and
\begin{equation}
\label{avtebv}
A_{V_T}/E(B-V)=-0.2025C^3+0.5256C^2-0.091C+3.49\,,
\end{equation}
where $C$ is $(B_T-V_T)_0$.
It is seen from these equations that for $0.8<(B_T-V_T)_0<1.9$~mag, $E(B_T-V_T)$ exceeds $E(B-V)$ by less than 15 per cent.
The calculation of $(B_T-V_T)_0$, $E(B_T-V_T)$ and $A_{V_T}$ by use of the equations~(\ref{dered}), (\ref{ebtvtebv}) and (\ref{avtebv}) 
needs several iterations.
We do not consider any spatial variations of the extinction law because even with $2.8<R_V<4.0$ for the space under consideration \citep{rv},
these variations negligibly affect the results.

\begin{figure*}
\includegraphics{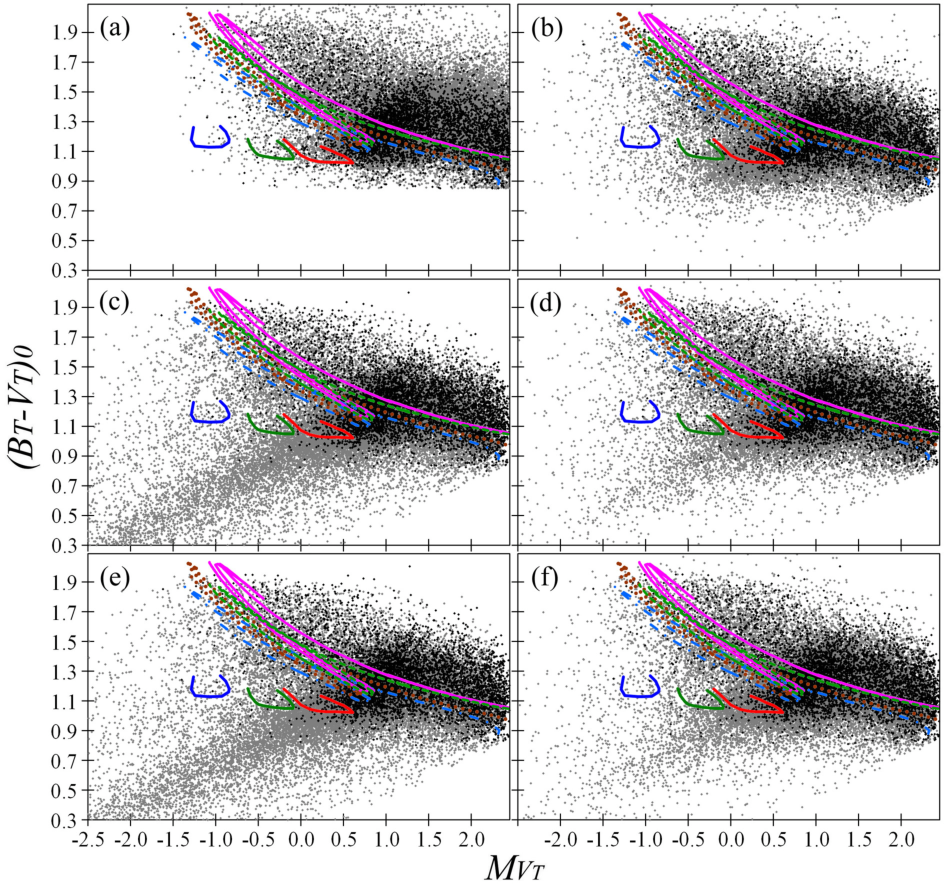}
\caption{HR diagram $M_{V_T}$ versus $(B_T-V_T)_0$ for the giants under consideration with $|b|<50\degr$ -- grey dots, $|b|>50\degr$ -- black dots.
The data are corrected for reddening and extinction from the sources:
(a) zero extinction,
(b) \citetalias{arenou},
(c) \citetalias{sfd},
(d) SFD$_R$,
(e) \citetalias{2015ApJ...798...88M}, 
(f) PLA$_R$,
(g) \citetalias{drimmel},
(h) \citetalias{green} (only for stars with valid estimates),
(i) \citetalias{av},
(j) \citetalias{g17},
(k) \citetalias{chen} and
(l) \citetalias{sale}.
The PARSEC isochrones at the young clump for $\mathbf Z=0.0152$ and age 0.25, 0.40, 0.63 Gyr -- blue, green and red solid curves, respectively.
The isochrones at the branch, clump and asymptotic branch:
2 Gyr, $[[Fe/H]]=-0.10$ PARSEC -- light blue dash-dotted, MIST -- brown dotted, 
5 Gyr, $[[Fe/H]]=-0.14$ PARSEC -- green dashed, MIST -- purple solid curve.
}
\label{hrdots1}
\end{figure*}

\begin{figure*}
\includegraphics{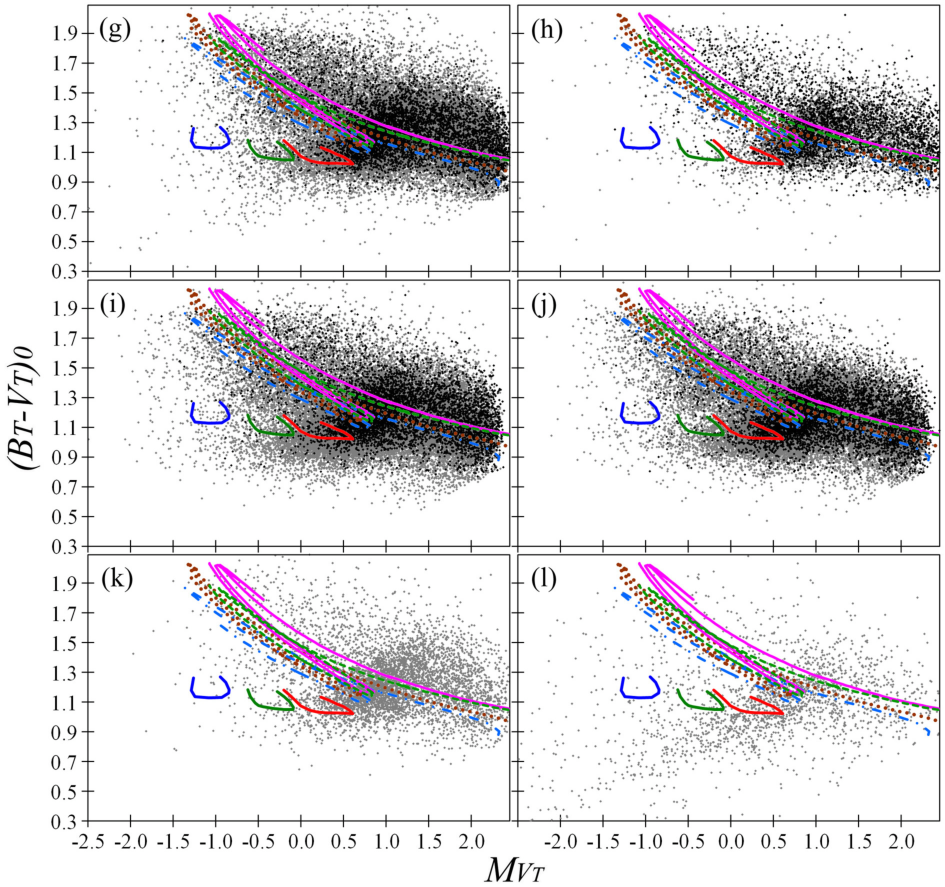}
\contcaption{
}
\label{hrdots2}
\end{figure*}

A detailed description of main kinds of the giants is given by \citet{clump}.
Fig.~\ref{hrdots1} shows these kinds: 
the fainter part of the branch -- the rightmost bulk at $M_{V_T}>1.2$ mag,
the main clump at $0.5<M_{V_T}<1.2$ mag and the separation of the most luminous giants at $M_{V_T}<0.5$ mag into 
the brighter branch and the young clump
\footnote{\citet{clump} refer to this kind as the vertical structure, but we prefer `young clump'}, 
respectively, redder and bluer than $(B_T-V_T)_0\approx1.2$ mag.

The parts of the isochrones at the branch, clump, and asymptotic branch are shown for the comparison:
for 2 Gyr ($[[Fe/H]]=-0.1$) and 5 Gyr ($[Fe/H]=-0.14$)
for PARSEC -- the light-blue dash-dotted and green dashed curves, and MIST -- the brown dotted and purple solid curves, respectively.
The solar metallicity $\mathbf Z=0.0152$ accepted in PARSEC from \citet{bressan} is comparable with the protosolar metallicity $\mathbf Z=0.0142$ 
accepted in MIST from \citet{asplund}.
The relation `age versus metallicity' for these isochrones is accepted on the basis of the TRILEGAL data for the sample, as discussed later.
It is seen that MIST gives slightly redder isochrones than PARSEC.

Any reasonable reddening and extinction would not mix the young clump, main clump, brighter branch, and fainter branch.
Yet, it is evident that the different estimates of reddening/extinction are responsible for the main difference of the plots of Fig.~\ref{hrdots1}.
The bulk of the stars is shifted to the right/up with lower extinction/reddening or to the left/down with higher ones.
It is especially evident in plots (c), (d), (e), (f), and (l) with the estimates for $|b|<50\degr$ from \citetalias{sfd}, SFD$_R$, \citetalias{2015ApJ...798...88M}, PLA$_R$, and \citetalias{sale}, respectively:
plenty of giants with strongly overestimated reddening/extinction form a narrow grey bulk from the centre to the left-down corner.
Initially, our sample keeps a certain level of completeness everywhere in the considered part of the HR diagram.
However, the giants with considerably overestimated reddening/extinction migrate from the main clump and fainter branch making them very incomplete
(for example, \citetalias{sfd} and \citetalias{2015ApJ...798...88M} lose more than 20 per cent of giants at the main clump and fainter branch). 
Moreover, in this way main clump and fainter branch contaminate the young clump.
As a result, any further analysis of the distribution of the giants with the estimates for $|b|<50\degr$ from \citetalias{sfd}, SFD$_R$, \citetalias{2015ApJ...798...88M}, PLA$_R$, and \citetalias{sale} is
strongly biased and makes little sense.
However, the distribution of the black dots in Fig.~\ref{hrdots1} shows that for $|b|>50\degr$ these data sources do not provide such a strong 
overestimation of the reddening and, thus, can provide useful results.

For $M_{V_T}<0.5$ mag, we do not consider the brighter branch giants due to their various and poorly defined ages and metallicities.
Also, due to a contamination of the young clump by the main clump and brighter branch, their $(B_T-V_T)_0$ has a large spread.
However, the young clump giants dominate among the bluest giants. 
Being younger than 2 Gyr, they have a well-defined average metallicity. 
It is nearly equal to the solar metallicity, and lies certainly within $-0.06<[Fe/H]<0.05$, as follows from TRILEGAL in agreement with \citet{haywood}.
Moreover, their $mode((B_T-V_T)_0)$ is almost independent of age, 
as evident from the young clump isochrones.
Such PARSEC isochrones are shown in Fig.~\ref{hrdots1} by the blue, green and red solid curves at the centres of the plots, for 0.25, 0.40, and 0.63 Gyr, respectively,
and $\mathbf Z=0.0152$.
The lowest parts of these curves are the domains of the slowest evolution of the giants, thus, constituting the young clump.
A lower envelope of such isochrones must fit the $mode((B_T-V_T)_0)$ of any real young clump sample, if
the reddening estimates are correct.
Indeed, the young clump looks like a horizontal bulk of stars along this lower envelope of the isochrones
in Fig.~\ref{hrdots1}~(a), (b), (g), (h), (i), (j), and (k) for the zero extinction, \citetalias{arenou}, \citetalias{drimmel}, \citetalias{green}, \citetalias{av}, \citetalias{g17}, and \citetalias{chen}, respectively.
A contamination of the young clump by the main clump and brighter branch can make $mode((B_T-V_T)_0)$ of the real sample slightly redder than this
lower envelope of the isochrones.
Yet, TRILEGAL, PARSEC and MIST suggest that somewhere within $-0.5<M_{V_T}<0.5$~mag this contamination is so negligible that the empirical
$\min(mode((B_T-V_T)_0))$ must always fit the lower envelope of the isochrones.
\footnote{$M_{V_T}$, which corresponds to $\min(mode((B_T-V_T)_0))$, is defined by a dominant age of the young clump stars. However, it is not important
because their $mode((B_T-V_T)_0)$ is almost independent of age.}
In case of systematic overestimation/underestimation of the reddening, the $\min(mode((B_T-V_T)_0))$ of a real young clump sample is 
bluer/redder than this lower envelope of the isochrones.
Thus, this can be a test of the reddening/extinction estimates for the young clump.
Obviously, the young clump is much bluer than this lower envelope in the above mentioned cases of strong overestimation of the reddening by 
\citetalias{sfd}, SFD$_R$, \citetalias{2015ApJ...798...88M}, PLA$_R$, and \citetalias{sale}.

For $M_{V_T}>0.5$ mag, the sample is quite homogeneous. It has an almost Gaussian distribution by $(B_T-V_T)_0$. 
Consequently, the average, median and mode of $(B_T-V_T)_0$ almost coincide. 
Hereafter, we consider the median $(B_T-V_T)_0$ as a function of $M_{V_T}$, separately
for $|b|<50\degr$ (31841 giants) and $|b|>50\degr$ (6233 giants) to reveal its variations with $|b|$.
Since the vast majority of the giants have $|b|<50\degr$, the results for all giants almost coincide with the ones for $|b|<50\degr$.

\begin{figure}
\includegraphics{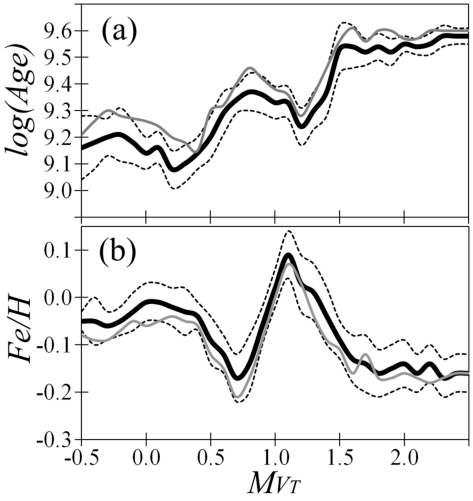}
\caption{TRILEGAL (a) $\log$(Age) versus $M_{V_T}$ and (b) $[Fe/H]$ versus $M_{V_T}$ diagrams for the giants under consideration:
average for $|b|<50\degr$ -- thick black curve, average for $|b|>50\degr$ -- thin grey curve,
the range of the results for $|b|<50\degr$ due to any reasonable variations of the TRILEGAL parameters -- dashed curves.
}
\label{tri}
\end{figure}

The isochrones are not enough to determine the theoretical position of the sample in the HR diagram because of some variations of the average
age and metallicity of the sample over the diagram.
We calculate the theoretical age and metallicity as some functions
of $M_{V_T}$. For this, we use TRILEGAL with the same constraints as for our sample.
The calculations with the TRILEGAL default parameters from \citet{trilegal}, but the Sun at 8 kpc from the Galactic Centre and 
13 pc above the Galactic mid-plane \citep{rcg, rgb},
give results shown in Fig.~\ref{tri} by the solid black and grey curves, respectively, for $|b|<50\degr$ and $|b|>50\degr$.
The most important TRILEGAL default parameters are:
initial mass function (IMF) from \citet{chabrier}; binary fraction is 0.3 with mass ratios between 0.7 and 1; 
thin disc with squared hyperbolic secant variations of density along $Z$, exponential variations along $R$, local calibration 55.4082 M$_{\sun}$ pc$^{-2}$,
two-step star formation rate (SFR), age versus metallicity relation (AMR) from \citet{rochapinto} with $\alpha$-enhancement as suggested by \citet{fuhrmann};
as well as certain definitions of thick disc, halo and bulge.
The calculations have been repeated with the parameters, which were widely varied following \citet{trilegal}:
different IMF, SFR, properties of thin disc, thick disc and halo, no or much more binaries, various extinction estimates or no extinction at all, 
the Sun at 7 kpc from the Galactic Centre and up to 30 pc above the Galactic mid-plane, etc..
The ranges of the results for $|b|<50\degr$ are shown in Fig.~\ref{tri} by the dashed curves.

\begin{figure*}
\includegraphics{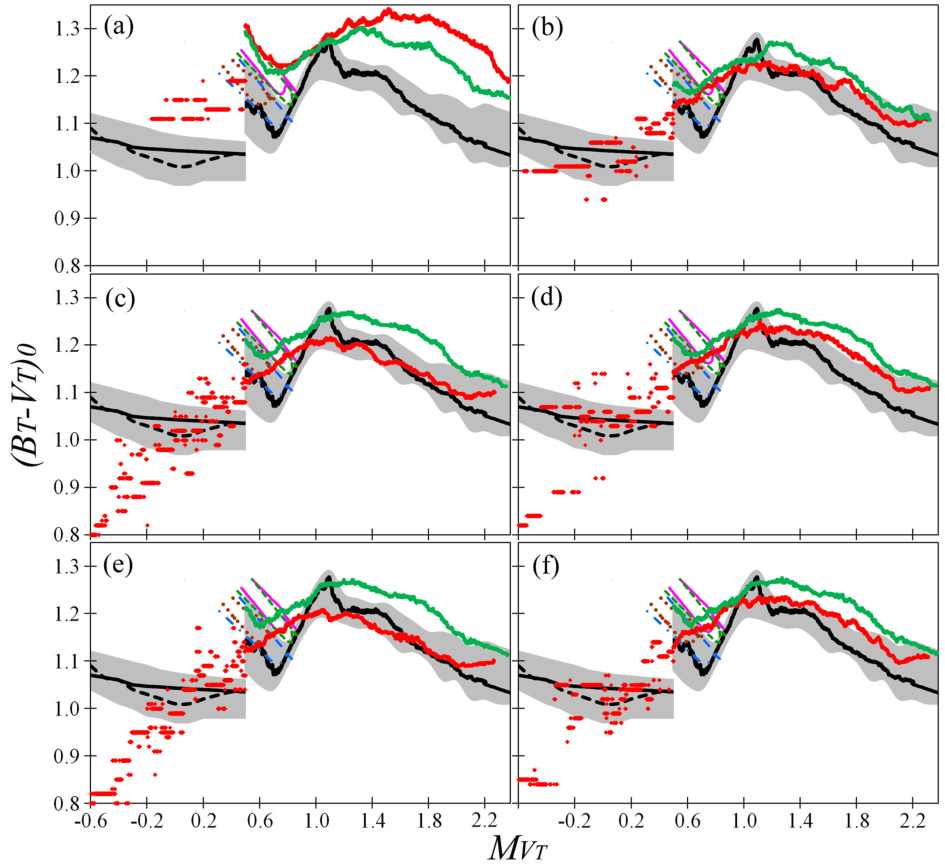}
\caption{HR diagram $M_{V_T}$ versus $(B_T-V_T)_0$. The young clump ($M_{V_T}<0.5$~mag):
$mode((B_T-V_T)_0)$ for the giants under consideration with $|b|<50\degr$ -- the red diamonds,
the lower envelope of the isochrones for $\mathbf Z=0.0152$ and age less than 0.63 Gyr from PARSEC -- the thin dashed black curve, 
MIST -- the thin solid black curve, with their range due to the variations of the average metallicity within $-0.06<[Fe/H]<0.05$ shown as the grey belt around it.
The fainter branch and main clump ($M_{V_T}>0.5$~mag):
the moving median $(B_T-V_T)_0$ for the giants under consideration with $|b|<50\degr$ -- the thick red curve,
with $|b|>50\degr$ -- the thick green curve,
the same from TRILEGAL for $|b|<50\degr$ -- the thick black curve, with its range due to the variations of the TRILEGAL parameters 
shown as the grey belt around it.
The isochrones at the main clump are:
2 Gyr, $[Fe/H]=-0.10$ PARSEC -- the light blue dash-dotted, MIST -- the brown dotted, 
5 Gyr, $[Fe/H]=-0.14$ PARSEC -- the green dashed, MIST -- the purple solid curve.
The empirical data are corrected for reddening and extinction from the sources:
(a) the zero extinction,
(b) \citetalias{arenou},
(c) \citetalias{sfd},
(d) SFD$_R$,
(e) \citetalias{2015ApJ...798...88M}, 
(f) PLA$_R$,
(g) \citetalias{drimmel},
(h) \citetalias{green} (only for stars with valid estimates),
(i) \citetalias{av},
(j) \citetalias{g17},
(k) \citetalias{chen} and
(l) \citetalias{sale}.
}
\label{hr1}
\end{figure*}

\begin{figure*}
\includegraphics{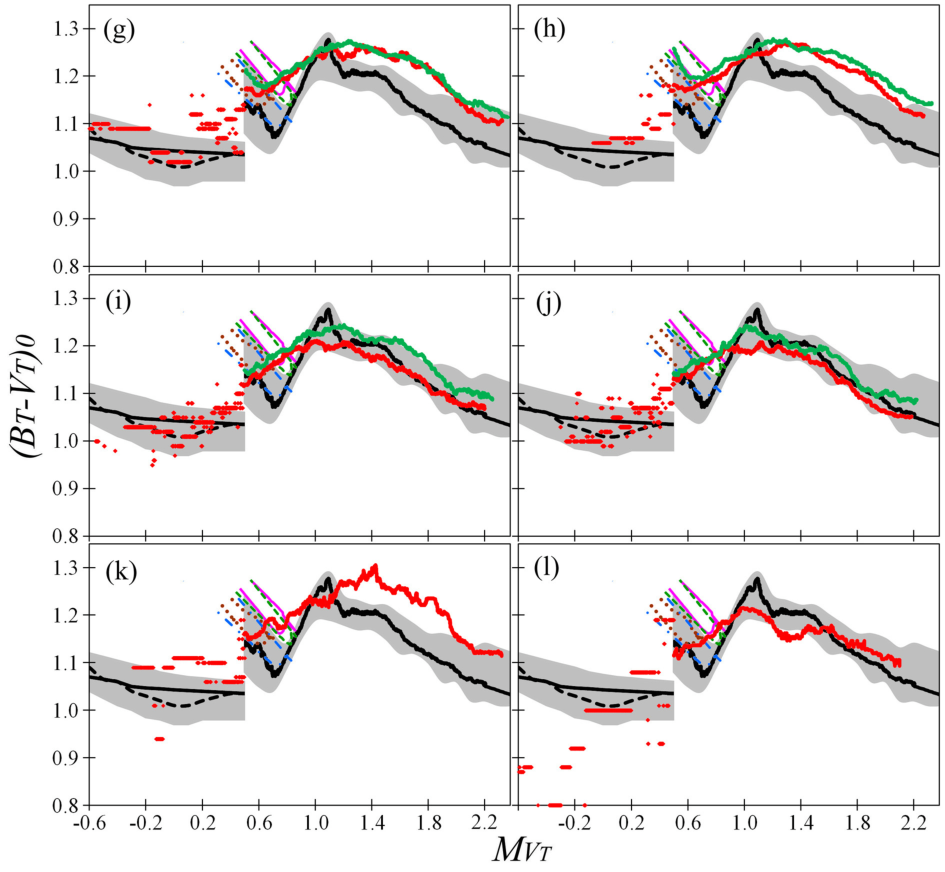}
\contcaption{
}
\label{hr2}
\end{figure*}

For $M_{V_T}>0.5$ mag and $|b|<50\degr$, the calculated theoretical median $(B_T-V_T)_0$ as a function of $M_{V_T}$ 
is shown in Fig.~\ref{hr1} by the thick solid black curve,
with its range due to the variations of the TRILEGAL parameters shown as the grey belt around it.
This belt also includes the uncertainties of the average age and metallicity shown in Fig.~\ref{tri}.
Such a relation `$(B_T-V_T)_0$ versus $M_{V_T}$' for $|b|>50\degr$ almost coincides with the one for $|b|<50\degr$ because, 
as evident from Fig.~\ref{tri},
the giants at $|b|>50\degr$ have a higher age (making the giants redder) but lower metallicity (making them bluer), 
which compensate each other in the relation `$(B_T-V_T)_0$ versus $M_{V_T}$'.
Therefore, the relation for $|b|>50\degr$ is not shown in Fig.~\ref{hr1}.

For $M_{V_T}<0.5$ mag, the theoretical lower envelope of the above mentioned young clump isochrones from PARSEC and MIST 
is shown in Fig.~\ref{hr1} by the dashed and solid black curves, respectively.
The grey belt around it shows the variations of the relation due to the accepted average metallicity within $-0.06<[Fe/H]<0.05$.

The data for the real giants under consideration are shown in Fig.~\ref{hr1} as well.
Different plots in Fig.~\ref{hr1} show the same reddening/extinction estimates used for the stars, as in Fig.~\ref{hrdots1}.
For $M_{V_T}>0.5$ mag,
the moving median $(B_T-V_T)_0$ over 1000 points for $|b|<50\degr$ and 500 points for $|b|>50\degr$ is shown by the thick red and green curves,
respectively.
For \citetalias{green}, \citetalias{chen}, and \citetalias{sale}, 500 and 200 points are used instead of 1000 and 500, respectively, for the moving windows.
For $M_{V_T}<0.5$ mag, the moving $mode((B_T-V_T)_0)$ over 500 points for $|b|<50\degr$ is shown by the red diamonds.
Being rather young, the young clump giants are rare at $|b|>50\degr$.

If the TRILEGAL model with some reasonable parameters of the Galaxy is correct 
and if the solar metallicity is $\mathbf Z\approx0.0152$ (we discuss this later),
then any correct estimates of reddening/extinction would put in Fig.~\ref{hr1}

\begin{enumerate}
\item the thick red curve near the thick black one, within the grey belt -- this is the case for the plots (c), (e), (i) and (j) 
for \citetalias{sfd}, \citetalias{2015ApJ...798...88M}, \citetalias{av}, and \citetalias{g17}, respectively
(the disagreement for all the plots at $0.8<M_{V_T}<0.9$~mag is, probably, due to an imperfection of TRILEGAL, as discussed later);
\item the thick green curve into the grey belt -- this is the case for the plots (i) and (j) for \citetalias{av} and \citetalias{g17}, respectively;
\item the lowest red diamonds near the thin black curves, within the grey belt -- 
this is the case for the plots (g), (h) and (j) for \citetalias{drimmel}, \citetalias{green}, and \citetalias{g17}, respectively.
\end{enumerate}
Also, the red curve almost fits the black thick curve within the grey belt for the plots (b), (d), (f), and (l) for \citetalias{arenou}, SFD$_R$, PLA$_R$, and \citetalias{sale}, respectively.
Yet, we have seen that for all of them, except \citetalias{arenou}, this is a strongly biased result.

%¹¹¹¹¹¹¹¹¹¹¹¹¹¹¹¹¹¹¹¹¹¹¹¹¹¹¹¹¹
\begin{table}
\def\baselinestretch{1}\normalsize\normalsize
\caption[]{Statistic tests for the main clump and fainter branch.
}
\label{test}
\[
\begin{tabular}{lcccc}
\hline
\noalign{\smallskip}
Source  & \multicolumn{2}{c}{$\mathcal{D}$} & \multicolumn{2}{c}{D-value} \\
\hline
\noalign{\smallskip}
 & $|b|<50\degr$ & $|b|>50\degr$ & $|b|<50\degr$ & $|b|>50\degr$ \\
\hline
\noalign{\smallskip}
Zero  & 3.513 & 2.662 & 0.370 & 0.276 \\
\citetalias{arenou} & 1.184 & 1.617 & 0.191 & 0.218 \\
\citetalias{sfd} & 1.109 & 1.847 & 0.167 & 0.231 \\
SFD$_R$ & 1.292 & 1.930 & 0.206 & 0.233 \\
\citetalias{2015ApJ...798...88M} & 1.105 & 1.854 & 0.165 & 0.230 \\
PLA$_R$ & 1.226 & 1.938 & 0.200 & 0.233 \\
\citetalias{drimmel} & 1.731 & 1.921 & 0.243 & 0.232 \\
\citetalias{green} & 1.869 & 2.091 & 0.229 & 0.248 \\
\citetalias{av} & 1.131 & 1.216 & 0.171 & 0.196 \\
\citetalias{g17} & 1.132 & 1.170 & 0.160 & 0.182 \\
\citetalias{chen} & 2.066 & & 0.250 & \\
\citetalias{sale} & 1.168 & & 0.164 & \\
\hline
\end{tabular}
\]
\end{table}

%¹¹¹¹¹¹¹¹¹¹¹¹¹¹¹¹¹¹¹¹¹¹¹¹¹¹¹¹¹¹¹¹¹¹¹¹

To test the disagreement of the empirical data with TRILEGAL at $0.8<M_{V_T}<0.9$~mag, we show in Fig.~\ref{hr1}
the main clump parts of the above mentioned isochrones. 
It is seen that the empirical data better fit the MIST (the brown dotted and purple solid curves) than PARSEC (the light blue dash-dotted and green dashed curves) 
isochrones.
This may point out to an imperfection of TRILEGAL, because it is based on PARSEC.

One can see several cases when the empirical results agree (or almost agree) with the theoretical ones for the fainter branch and disagree for 
the young clump (\citetalias{arenou}, \citetalias{sfd}, SFD$_R$, \citetalias{2015ApJ...798...88M}, PLA$_R$, \citetalias{av} and \citetalias{sale}), 
and vice versa (\citetalias{drimmel} and \citetalias{green}).
Due to a correlation between $R$ and $M_{V_T}$, the young clump giants have a slightly larger median $R$ (350 pc) than for the fainter branch (300 pc).
Yet, the difference is small.
It can explain the cases of (\citetalias{arenou}, \citetalias{av})/(\citetalias{drimmel}, \citetalias{green}) as an overestimation/underestimation of the reddening for the most distant/nearby giants under consideration
due to the adjustment of these data sources to the nearby/distant stars only.
However, remaining such the cases are solely explained by the migration of the main clump and fainter branch giants to the domain of
the young clump due to the overestimated reddening/extinction.

We perform two tests to estimate the agreement of the empirical and theoretical relations `$(B_T-V_T)_0$ versus $M_{V_T}$' for the main clump and fainter 
branch. The first test is the 2D Kolmogorov-Smirnov (K--S) test which determines if two data sets differ significantly. Here, we compare the empirical set of points for 
each map in the space `$(B_T-V_T)_0$ versus $M_{V_T}$' with the randomly selected points from all the variety of the used TRILIGAL models. 
We present the D-value (or K--S statistic) which, by definition, is the absolute maximal distance between the cumulative distribution functions of the two samples. 
In other words, the closer this number is to 0, the more likely it is that the two samples were drawn from the same distribution.
The second test is similar to the first one but it deals with a median average curve for empirical data and the model spread which is presented by the grey belt 
in Fig.~\ref{hr1}. 
For each extinction/reddening source, we compute the following value:
\begin{equation}
\label{Av_dist}
\mathcal{D} = \frac{1}{N} \sum_{i=1}^{N} \frac{|y_i-y_{b,i}| + |y_i-y_{t,i}|}{|y_{t,i}-y_{b,i}|}\,,
\end{equation}
where $N$ is the total number of points in the median average curve, $y_i$ is the ordinate of the median average curve at the abscissa $x_i$ in the space 
`$(B_T-V_T)_0$ versus $M_{V_T}$', $y_{b,i}$ is an interpolation of the bottom envelope of the model spread at 
$x_i$ and $y_{t,i}$ is an interpolation of the top envelope of the model spread at $x_i$. 
As such, $\mathcal{D}$ is an averaged distance between the empirical data and the model spread.
   
The results are presented in Table~\ref{test}. It shows that the best agreement is for \citetalias{av} and \citetalias{g17}
(formally, \citetalias{sfd}, \citetalias{2015ApJ...798...88M}, and \citetalias{sale} also give fine test results, but they have been criticized because of the strong biases).

For the young clump, the theoretical relation `$(B_T-V_T)_0$ versus $M_{V_T}$' is presented by the lower envelope of the isochrones, but not by a
TRILEGAL simulated sample. Therefore, for the young clump the only possible test is simple: 
does $\min(mode((B_T-V_T)_0))$ fit the constraint $-0.06<[Fe/H]<0.05$, i.e. does the lowest red diamond lie within the grey belt?
As seen from Fig.~\ref{hr1}, only \citetalias{drimmel}, \citetalias{green}, and \citetalias{g17} pass this test.

\section{Discussion}

%¹¹¹¹¹¹¹¹¹¹¹¹¹¹¹¹¹¹¹¹¹¹¹¹¹¹¹¹¹
\begin{table}
\def\baselinestretch{1}\normalsize\normalsize
\caption[]{Median $E(B_T-V_T)$ at $|b|>50\degr$.
}
\label{deg50}
\[
\begin{tabular}{lccc}
\hline
\noalign{\smallskip}
 Source & Median $E(B_T-V_T)$ & $\Delta E(B_T-V_T)$ & $\sum E(B_T-V_T)$ \\
\hline
\noalign{\smallskip}
Zero & 0.000 & 0.070 & 0.070 \\
\citetalias{arenou} & 0.035 & 0.030 & 0.065 \\
\citetalias{sfd} & 0.023 & 0.045 & 0.068 \\
SFD$_R$ & 0.021 & 0.045 & 0.066 \\
\citetalias{2015ApJ...798...88M} & 0.021 & 0.045 & 0.066 \\
PLA$_R$ & 0.019 & 0.045 & 0.064 \\
\citetalias{drimmel} & 0.021 & 0.045 & 0.066 \\
\citetalias{green} & 0.017 & 0.050 & 0.067 \\
\citetalias{av} & 0.053 & 0.010 & 0.063 \\
\citetalias{g17} & 0.068 & 0.000 & 0.068 \\
\hline
\end{tabular}
\]
\end{table}

%¹¹¹¹¹¹¹¹¹¹¹¹¹¹¹¹¹¹¹¹¹¹¹¹¹¹¹¹¹¹¹¹¹¹¹¹

In Fig.~\ref{hr1}, the red curves for \citetalias{drimmel}, \citetalias{green}, and \citetalias{chen} are not strongly biased, yet, they are located far from the grey belt. 
This means that these data sources generally underestimate the reddening for the giants within $|b|<50\degr$.

However, the most important conclusions can be made for $|b|>50\degr$, 
where all the data sources under consideration seem to give the unbiased results.
As noted earlier, only the estimates of \citetalias{av} and \citetalias{g17} fit TRILEGAL (the green curve is inside the grey belt).
Table~\ref{deg50} contains some estimates for all the data sources at $|b|>50\degr$.
They are the median $E(B_T-V_T)$, the minimal value $\Delta E(B_T-V_T)$ of the reddening underestimation taken from the shift of the the faint branch giants 
(green curve) off the TRILEGAL prediction (the grey belt) along the reddening/extinction line (i.e. the value needed to put the green curve inside the grey belt),
and the resulting sum $\sum E(B_T-V_T)$ of the median $E(B_T-V_T)$ and $\Delta E(B_T-V_T)$. 
The $\sum E(B_T-V_T)$ can be considered as a median reddening estimate for $|b|>50\degr$ by the data source and TRILEGAL.
Table~\ref{deg50} shows that \citetalias{sfd}, SFD$_R$, \citetalias{2015ApJ...798...88M}, PLA$_R$, \citetalias{drimmel}, and \citetalias{green} give the lowest median $E(B_T-V_T)$, \citetalias{av} and \citetalias{g17} -- the highest value, 
whereas \citetalias{arenou} is in-between
\footnote{\citetalias{arenou} accepted the single value $E(B-V)=0.032$~mag at $|b|>60\degr$ behind the dust layer ($R>300$~pc)
due to the lack of precise measurements of the reddening/extinction at high latitudes during the construction of \citetalias{arenou}.}
Yet, naturally, $\sum E(B_T-V_T)$ is almost the same for all the data sources.
This leads to the conclusion that the median $E(B_T-V_T)$ at $|b|>50\degr$ is indeed $0.066$~mag 
[the median $E(B-V)$ is $0.06$~mag following the equation~(\ref{ebtvtebv})] 
and, thus, a value $\Delta E(B_T-V_T)\approx0.045$~mag ($\Delta E(B-V)\approx0.041$~mag)
should be added to correct the median $E(B_T-V_T)$ from \citetalias{sfd}, \citetalias{2015ApJ...798...88M}, \citetalias{drimmel}, and \citetalias{green}.
Since the majority of the giants under consideration at $|b|>50\degr$ are behind the dust layer, this correction is also needed for the \citetalias{sfd} and \citetalias{2015ApJ...798...88M} 
reddening estimates to infinity.
To show this, we consider the traditionally focused values $E(B-V)$ in the Galactic polar caps, for example $15\degr$ around the northern (NGP) and 
southern (SGP) Galactic poles (a review of the reddening estimates at the poles is given, for example in the \citetalias{sfd} paper). 
We select the giants behind the dust layer with $|Z|>250$ pc.
Table~\ref{ngpsgp} shows that, again, \citetalias{sfd}, SFD$_R$, \citetalias{2015ApJ...798...88M}, PLA$_R$, \citetalias{drimmel}, and \citetalias{green} give the lowest estimates of the median $E(B-V)$, 
\citetalias{av} and \citetalias{g17} -- the highest, whereas \citetalias{arenou} is in-between. 
Yet, TRILEGAL supports that the median $E(B-V)$ is equal to 0.06~mag at the Galactic poles.

%¹¹¹¹¹¹¹¹¹¹¹¹¹¹¹¹¹¹¹¹¹¹¹¹¹¹¹¹¹
\begin{table}
\def\baselinestretch{1}\normalsize\normalsize
\caption[]{Median $E(B-V)$ within $15\degr$ from the Galactic poles for the giants with $|Z|>250$ pc.}
\label{ngpsgp}
\[
\begin{tabular}{lcc}
\hline
\noalign{\smallskip}
 Source & NGP & SGP \\
\hline
\noalign{\smallskip}
\citetalias{arenou} & 0.032 & 0.030 \\
\citetalias{sfd} & 0.018 & 0.017 \\
SFD$_R$ & 0.017 & 0.017 \\
\citetalias{2015ApJ...798...88M} & 0.013 & 0.016 \\
PLA$_R$ & 0.013 & 0.016 \\
\citetalias{drimmel} & 0.018 & 0.018 \\
\citetalias{green} & 0.015 & 0.013 \\
\citetalias{av} & 0.056 & 0.053 \\
\citetalias{g17} & 0.059 & 0.060 \\
\hline
\end{tabular}
\]
\end{table}

%¹¹¹¹¹¹¹¹¹¹¹¹¹¹¹¹¹¹¹¹¹¹¹¹¹¹¹¹¹¹¹¹¹¹¹¹

The agreement or disagreement of the reddening data sources with each other is seen from their linear correlation coefficients.
For $|b|>50\degr$ they are high ($>0.73$) only for all the pairs of \citetalias{sfd}, SFD$_R$, \citetalias{2015ApJ...798...88M}, PLA$_R$, \citetalias{drimmel}, and \citetalias{green}. 
Also these correlation coefficients show that \citetalias{drimmel} is closer to \citetalias{sfd}, whereas \citetalias{green} is closer to \citetalias{2015ApJ...798...88M}.
For all the giants under consideration the correlation coefficients are high for all the pairs of \citetalias{sfd}, SFD$_R$, \citetalias{2015ApJ...798...88M}, and PLA$_R$ ($>0.69$);
for all the pairs of \citetalias{arenou}, \citetalias{av}, and \citetalias{g17} ($>0.72$), as well as for \citetalias{green} versus \citetalias{g17} (0.65), \citetalias{green} versus \citetalias{av} (0.55), \citetalias{green} versus \citetalias{arenou} (0.59)
and \citetalias{green} versus \citetalias{drimmel} (0.51). For the rest pairs, the correlation coefficients are lower.

This means that at $|b|<50\degr$ some compromised reddening estimates can be found from the highly correlated estimates of 
\citetalias{g17} and \citetalias{green} (remembering that \citetalias{arenou} is based on very limited data, whereas \citetalias{av} is a sister of \citetalias{g17} by its origin).
However, at $|b|>50\degr$ there is no such a compromise: one has to choose either a lower reddening from \citetalias{sfd}, \citetalias{2015ApJ...798...88M}, \citetalias{drimmel}, and \citetalias{green} or a higher reddening from
\citetalias{g17}.

Besides the underestimated reddening at high latitudes, there are two other explanations for the discrepancy of the green curves and the grey belt 
in Fig.~\ref{hr1}:
\begin{enumerate}
\item TRILEGAL has not considered a reliable set of the parameters of the Galaxy, and
\item the solar metallicity is $\mathbf Z>0.021$ instead of $0.0152$: with a fixed $[Fe/H]$ in TRILEGAL it gives a much higher $\mathbf Z$ for the giants 
under consideration and changes the TRILEGAL's `$(B_T-V_T)_0$ versus $M_{V_T}$' relation greatly.
\end{enumerate}
We regard the former alternative as improbable. 
For the latter alternative our approach sets an almost linear relation between the accepted solar metallicity and median reddening at high latitudes:
$\mathbf Z=0.015$ corresponds to $E(B-V)=0.06$, whereas $\mathbf Z=0.021$ to $E(B-V)=0.02$, with, respectively,
\citetalias{g17} or \citetalias{sfd}, SFD$_R$, \citetalias{2015ApJ...798...88M}, PLA$_R$, \citetalias{drimmel} and \citetalias{green} as the best estimates.
A compromise with $\mathbf Z=0.018$ and median $E(B-V)=0.04$ is also possible.
We note that the same is applicable to the previous study of the O--F stars \citep{gm2017big}:
the same increase of the accepted solar metallicity would shift the isochrones and make the \citetalias{drimmel} estimates the best.

Yet, there have been some robust arguments to accept the solar metallicity as $\mathbf Z=0.0152$ in PARSEC \citep{bressan}
or even $\mathbf Z=0.0142$ in MIST \citep{asplund}.
Therefore, we should explain the suggested underestimation of the reddening at $|b|>50\degr$ by \citetalias{sfd}, SFD$_R$, \citetalias{2015ApJ...798...88M}, PLA$_R$, \citetalias{drimmel} and \citetalias{green}.

\citetalias{drimmel} and \citetalias{green} are constructed with two `boundary conditions': the zero reddening at the Sun, and the \citetalias{sfd} reddening behind the dust layer.
Therefore, any systematic error of \citetalias{sfd} would appear in \citetalias{drimmel} and \citetalias{green}.

Consequently, we should find an initial source of systematic errors, common for \citetalias{sfd} and \citetalias{2015ApJ...798...88M}.
For any emission-based 2D reddening map, such as \citetalias{sfd} or \citetalias{2015ApJ...798...88M}, the reddening through the dust half-layer is estimated from 
the emission generated in this half-layer.
The emission-to-reddening calibrations for \citetalias{sfd} and \citetalias{2015ApJ...798...88M} are based on some samples of reddened elliptical galaxies and quasars, respectively.
The same for both \citetalias{sfd} and \citetalias{2015ApJ...798...88M}, the cosmic IR background emission from the intergalactic dust and 
Zodiacal IR foreground emission from the dust in the Solar system
are removed at early stages of the data processing in order to obtain the emission from the Galactic dust.
As a result, the median $E(B-V)$ through the dust half-layer in the large low dust column density regions of the sky was set to $0.017\pm0.028$~mag in \citetalias{sfd} 
and to a similar very low value in \citetalias{2015ApJ...798...88M}. In fact, these values appear in Table~\ref{ngpsgp}, whereas these regions of the sky are tabulated in table~5 of the 
\citetalias{sfd} paper.
The uncertainty of this emission-to-reddening calibration is so high that the median $E(B-V)$ $0.017+0.028=0.045$~mag instead of $0.017$~mag is quite probable.
\citetalias{sfd} attempted to use counts of galaxies (after the reddened colours of the galaxies had been used)
to improve this calibration for low reddenings.
They obtained a value which was at least twice larger than that which had been obtained by use of the reddening of the galaxies.
Apart from other explanations, the most natural reason is that the real reddening at the regions with low reddening is indeed at least two times higher 
than presented in \citetalias{sfd}.

This systematic underestimation of low reddening may appear because it is hard to distinguish the emissions from the Galactic, intergalactic and 
Zodiacal dust in the sky areas with low reddening/emission.
The uncertainty of this distinguishing of the Galactic and Zodiacal emission may be enhanced by the orientation of the ecliptic with respect to the Galactic
equator. They intersect at a large angle and near the directions to the Galactic Centre and anticentre.
As a result, two parts of the Zodiacal dust belt are projected to the most dense parts of the dust layer of the Gould Belt at $l=345\degr$, $b=19\degr$ and 
$l=165\degr$, $b=-19\degr$ \citep[table 4]{av}, \citep[see also][]{gould, astroph}.
Other parts of the Zodiacal dust belt are projected to the high latitude areas
(for example, ecliptic at the longitudes $155\degr<\lambda<205\degr$ and $-25\degr<\lambda<25\degr$, i.e. in Leo, Virgo, Aquarius and Pisces, 
drops in the areas with $|b|>50\degr$).
In fact, in  both the cases we know only the total reddening: `Gould Belt plus Zodiacal belt' and `High latitudes plus Zodiacal belt'.
Underestimation of the reddening in the Gould Belt has been a common practice. It must lead to an overestimation of the Zodiacal dust reddening and,
consequently, to an underestimation of the reddening at high latitudes.
This may be the case for \citetalias{sfd} and \citetalias{2015ApJ...798...88M} and should be tested in future studies.

Moreover, by use of the emission generated by the dust grains of one size, \citetalias{sfd} and \citetalias{2015ApJ...798...88M} estimate the reddening generated by the grains of another size.
However, some strong variations of the extinction law far from the Galactic mid-plane, obviously, due to the variations of the distribution of grains on size 
have been discovered \citep[see e.g.][]{rv, g2013, g2016, astroph}.
These lead to some still poorly known spatial variations of the emission-to-reddening calibration.

Finally, we shall explain the advantages of \citetalias{g17} as, apparently, the best reddening estimates for the local stars.
\begin{enumerate}
\item Unlike \citetalias{sfd}, \citetalias{2015ApJ...798...88M}, \citetalias{drimmel}, \citetalias{green}, \citetalias{chen}, and \citetalias{sale}, \citetalias{g17} uses the local stars, including the ones within $R<415$ pc.
The turn-off stars are used knowingly in order to deal with a rather complete sample within $R<1200$ and $|Z|<600$ pc.
However, we emphasize that \citetalias{g17} would be far from the best and even unreliable for distant regions of the dust layer, i.e. for $R>1200$ with $|Z|<200$ pc.
\item \citetalias{g17} uses the photometry of millions of stars instead of tens of thousands for \citetalias{arenou}.
\item \citetalias{g17} is a much more detailed description of the local dust medium than \citetalias{av}.
\item Unlike the others, \citetalias{g17} combines the photometry and star counts to derive the reddening. The photometry is not used directly, but 
only to select the counted stars. This eliminates some systematic errors.
It is the most important in the space with low reddening, e.g. at high latitudes, where even a small error of, say, 0.04~mag in the accepted 
dereddened colour becomes a reddening systematic zero-point offset.
This approach allows us to obtain the estimates at high latitudes which are independent of \citetalias{sfd}, \citetalias{2015ApJ...798...88M} or any other emission-based map.
This eliminates any error of the emission-reddening calibration, which may affect \citetalias{drimmel}, \citetalias{green}, \citetalias{chen}, \citetalias{sale} and other \citetalias{sfd} followers.
\end{enumerate}

\section{Conclusions}

TGAS parallaxes, {\it Tycho-2} photometry and reddening/extinction estimates from nine data sources
for 38074 giants within 415 pc from the Sun have been used to compare their position in the HR diagram with the theoretical estimates based on 
the PARSEC and MIST isochrones and the TRILEGAL model of the Galaxy with its parameters being widely varied.
The accuracy of the data allows us to reveal some considerable systematic errors of the reddening/extinction estimates as the main uncertainty in 
the positioning of the giants in the HR diagram.

This study confirms the same conclusions for the giants as \citet{gm2017big} made for the O--F stars. 
For the solar metallicity $\mathbf Z<0.018$, the empirical data better fit the theoretical ones with the reddening/extinction estimates from 
recent 3D reddening map \citetalias{g17}. In this case, the median reddening at high Galactic latitudes behind the dust layer is $E(B-V)>0.04$~mag. 
Consequently, it has been considerably underestimated by the 2D reddening maps, which are
based on the dust emission observations by {\it IRAS}, {\it COBE}, and {\it Planck} (\citetalias{sfd} and \citetalias{2015ApJ...798...88M}), and by \citetalias{sfd}'s 3D followers (\citetalias{drimmel} and \citetalias{green}).
However, with a higher solar metallicity $\mathbf Z>0.018$ the reddening/extinction estimates from \citetalias{drimmel} and \citetalias{green} become the best.

Anyway, any emission-based 2D estimates of reddening, such as \citetalias{sfd} or \citetalias{2015ApJ...798...88M}, even reduced to a star's distance $R$ in assumption of the only flat layer of dust,
can be wrong due to a complex distribution of the dust within 415 pc.
This complexity of the local dust medium should not be described by too simplified models (\citetalias{drimmel} or \citetalias{arenou}) or without engaging photometry of the local stars
(\citetalias{green}, \citetalias{chen} and \citetalias{sale}).
Among the analytical models, \citetalias{av} is the best, describing the local dust medium by two intersected dust layers, 
at the Galactic mid-plane and in the Gould Belt.
Yet, the results of this study show that \citetalias{av} needs a further refinement.

\section*{Acknowledgements}

We thank an anonymous reviewer for useful comments.
We thank Jieun Choi and Aaron Dotter for the discussion of the MIST data.
AVM is partly supported by the Russian Foundation for Basic Researches (grant number 14-02-810).
AVM is a beneficiary of a mobility grant from the Belgian Federal Science Policy Office.
The resources of the Centre de Donn\'ees astronomiques de Strasbourg, Strasbourg, France 
(\url{http://cds.u-strasbg.fr}) including the reddening/extinction data sources
under consideration were widely used in this study.
This work has made use of data from the European Space Agency (ESA) mission {\it Hipparcos}.
This work has made use of data from the ESA mission {\it Gaia} (\url{https://www.cosmos.esa.int/gaia}), processed by
the {\it Gaia} Data Processing and Analysis Consortium (DPAC; \url{https://www.cosmos.esa.int/web/gaia/dpac/consortium}). Funding
for the DPAC has been provided by national institutions, in particular the institutions participating in the {\it Gaia} Multilateral Agreement.

\bsp	% typesetting comment
\label{lastpage}
\end{document}